\documentclass[epj,referee]{svjour}
\usepackage{graphics}
\usepackage{epsf}

  \def \bm #1{ \mbox {\boldmath $#1$} }

\begin{document}

\title{Polarization phenomena by deuteron fragmentation into pions}
\author{A.Yu.~Illarionov
 \and A.G.~Litvinenko \and G.I.~Lykasov
}                     
%
%
\institute{Joint Institute for Nuclear Research, 
           141980 Dubna, Moscow Region, Russia\\
\email{Alexei.Illarionov@jinr.ru}
}
%
\date{Received: \today }
%
\abstract{
  The fragmentation of deuterons into pions emitted forward in the
  kinematic region forbidden for free nucleon-nucleon collisions
  is analyzed. The inclusive relativistic invariant spectrum of pions and
  the tensor analyzing power $T_{20}$ are investigated within the framework
  of an impulse approximation using different kinds of the deuteron wave
  function. The influence of $P$-wave inclusion in the deuteron wave
  function is studied, too. The invariant spectrum is shown to be more
  sensitive to the amplitude of the $NN \to \pi X$ process than the tensor
  analyzing power $T_{20}$. 
It is shown that the inclusion of the non-nucleon degrees of freedom
in a deuteron results a satisfactory description of experimental data
about the inclusive pion spectrum and improves the description of 
data about $T_{20}$. According to the experimental data, $T_{20}$ has the
positive sign and very small values, less than $0.2$, what contradicts 
to the theoretical calculations ignoring these degrees of freedom. 
%
\PACS{
   {25.10.+s}{Nuclear reactions involving few-nucleon systems} \and
   {24.70.+s}{Polarization phenomena in reactions}   \and
   {24.10.Jv}{Relativistic models}
     } 
} 
\maketitle


\section{Introduction}
\label{intro}

 Among the main sources of information about the deuteron structure at
 small distances are the reactions of hadrons production by proton-deuteron
 and deuteron-deuteron collisions in the kinematic region forbidden for
 the free nucleon-nucleon interaction \cite{bal85,fra81}, the so-called
 cumulative processes. This kinematic region corresponds to the values of
 the light-cone variable
 $x = 2(E^\prime + p^\prime)/(E_D + p_D) \geq 1$, where $E^\prime, E_D$ and
 $p^\prime, p_D$ are the energies and momenta of the final hadron and
 deuteron, respectively.
 The nucleon momentum distributions in the deuteron, extracted from the
 reaction $D p \to p X$ at forward  proton emission and $eD$-inelastic
 scattering \cite{bos82} actually coincide with each other (see,
 for example, \cite{pun96}). So, one can conclude that hadron and
 lepton probes
 result in the same information about the deuteron structure. The so-called
 Paris \cite{lan81} and Reid \cite{rei68} deuteron wave functions (DWF)
 reproduce rather well the experimental data on the $D p \to p X$ reaction
 for internal momenta $k = \sqrt{m^2 / (4x(1 - x)) - m^2}$ up to $0.25$~GeV/c
 within the framework of an impulse approximation (IA) \cite{pun96}).
 The inclusion of corrections to IA related to secondary interactions
 allows one to describe the experimental data on the deuteron fragmentation
 $D p \to p X$ at $k > 0.25$~GeV/c \cite{lyk93}.

 The investigation of polarization phenomena by deuteron fragmentation
 at middle and high energies in the kinematic region forbidden for hadrons
 emission by free $N - N$ scattering has recently become very
 topical. Cumulative proton production in the collision of
 polarized deuterons with the target results in information about the
 deuteron spin structure at small inter-nuclear distances. This can be seen
 from the experimental and theoretical study of deuteron fragmentation into
 protons at a zero angle \cite{abl83,abl90,lyk93}. The theoretical analysis
 of this reaction has shown that the tensor analyzing power $T_{20}$ and
 polarization transfer coefficient $\kappa$ are more sensitive to the used
 deuteron wave function (DWF), particular to the reaction mechanism, than
 the inclusive spectrum \cite{lyk93}. At the present time,
 not one DWF relativistic
 form can describe $T_{20}$ at $x \ge 1.7$ measured by
 $D p \to p X$ stripping. On the other hand, the inclusion of the reaction
 mechanism: namely the impulse approximation and the secondary interaction of
 produced hadrons can describe both the inclusive spectrum and $T_{20}$ at
 $x \le 1.7$ using only the nucleon degrees of freedom \cite{lyk93}.
 One of the causes, to explain this phenomenon can be the fact
 that the deuteron structure at a high ($>~0.20$~GeV/c) internal momentum
 (short inter-nuclear distances $<~1$~fm) is determined by non-nucleon
 degrees of freedom. The inclusion of non-nucleon degrees of freedom,
 (it can be either the six-quark state or the composition of $\Delta\Delta$,
 $NN^\star$, $NN\pi$ and other states in the deuteron) allowed one
 to describe the
 experimental data on the inclusive proton spectrum at $x \geq 1.7$
 8\cite{lyk93}. A number of papers were dedicated to a theoretical
 analysis of the deuteron stripping to protons, see for example Refs. in
 \cite{fra81,lyk93}. However, to date there has been no unified theoretical
 description of $T_{20}$ on the whole kinematic region of protons emitted
 forward by the $D p \to p X$ stripping.

 If we try to study the manifestation of non-nucleon degrees of freedom,
 it is natural to investigate the cumulative production of different hadrons
 having different quark contents. Interesting experimental
 data on $T_{20}$ in the reaction $D p \to \pi X$ where the pion is
 emitted forward have been published recently \cite{afa98}. They show very
 small,
 approximately constant, value of the tensor analyzing power $T_{20}$ for
 the deuteron fragmentation of into pions $D p \to \pi X$ at $x \geq 1$. The
 mechanism of this reaction is mainly an impulse approximation as
 the secondary interaction or the final state interaction
 (FSI) is very small and can be neglected \cite{ame81}. Quite large yield
 of high momentum pions produced by $p - D$ and $p - A$ collisions in the
 kinematical region forbidden for free $N - N$ scattering was explained
 by both few-nucleon correlation models \cite{fra81} and \cite{efr88} or
 a multi-quark bags one \cite{luk79,bur84}. However, the polarization
 phenomena in deuteron fragmentation into pions were outside of these
 models.

 In this paper, we present a relativistic invariant analysis of the deuteron
 tensor analyzing power $T_{20}$ and unpolarized pion spectrum in the
 backward
 inclusive $p + D \to \pi(180^0) + X$ reaction (in deuteron rest frame).
 The main goal is to describe
 this reaction in a consistent relativistic approach using the nucleon
 model of the deuteron. A fully covariant expression for all quantities is
 obtained
 within the Bethe-Salpeter (BS) formalism. This way results in general
 conclusions of the process amplitude which can not be seen in the
 non-relativistic approach. On the other hand, the non-relativistic limit
 will be recovered, and some links to non-relativistic corrections can be
 found. This analysis of the deuteron models can be very important to
 search for nuclear quark phenomena.

 \section{Relativistic impulse approximation}
 \label{sec:ria}

\begin{figure}\sidecaption
\centerline{
\resizebox{0.35\textwidth}{!}{%
  \includegraphics{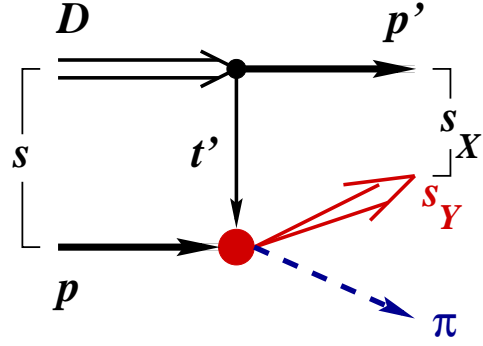}
}}
\caption{Relativistic impulse approximation diagram.}
\label{fig:ria}       
\end{figure}

 Let us consider the inclusive reaction of deuteron fragmentation to pion:
\begin{equation}
 {\vec D} + p \to \pi(0^0) + X
\label{first}
\end{equation}
 within the framework of the impulse approximation, fig.~\ref{fig:ria}.

 The amplitude of this process ${\cal T}_{pD}^\pi$ can be written in the
 following relativistic invariant form:
\begin{eqnarray}
 {\cal T}_{pD}^\pi \equiv {\cal T}(Dp \to \pi X) =
 \left(\bar{\cal U}_Y\Gamma_{NN}\right)_{\alpha\beta}
 \bar u^{(\sigma_{p'})}_\gamma(p')
\nonumber \\
 \times \left(\widehat n+m\over n^2-m^2\right)_{\beta\delta}
 u^{(\sigma_p)}_\alpha(p)\left(\Gamma_\mu(D,q){\cal C}\right)_{\delta\gamma}
 \xi^\mu_M(D)~.
\label{ND1}
\end{eqnarray}
 where $(\bar{\cal U}_Y\Gamma_{NN})$ is the truncated $NN \to \pi Y$ vertex;
 $\alpha, \beta, \gamma$ and $\delta$ are the Dirac indices (with summation
 over twofold indices); $\mu$ is a Lorentz index;
 ${\cal C} = i\gamma_2\gamma_0$ is a charge conjugation Dirac matrix and
 $M$ is the deuteron spin projection. Here, the
 deuteron vertex $\left(\Gamma_\mu(D,q){\cal C}\right)$  satisfies the $BS$
 equation and depends on the relative momentum $q=(n-p')/2$ and total
 momentum $D=n+p'$ of deuteron, $\xi^\mu_M(D)$ is the four-vector of
 the deuteron polarization. It satisfies the following equations:
 \begin{eqnarray}
 &&\xi^{\mu M}(D)D_\mu=0~,~~\xi^{\mu M}(D)\xi_{\mu M'}(D)=-\delta_{M'}^M
 \nonumber \\
 &&~~\sum_M \left(\xi_{\mu M}(D)\right)^*\xi_{\nu M}(D)=
 -g_{\mu\nu}+{D_\mu D_\nu\over M^2},
 \label{pol_vec}
 \end{eqnarray}
 Squaring this amplitude, one can write the relativistic invariant inclusive
 pion spectrum of the reaction $Dp \to \pi X$ in the following form:
 \begin{eqnarray}
 \rho_{pD}^\pi&=&\varepsilon_\pi{d\sigma\over d^3p_\pi} =
 {1 \over (2\pi)^3}
 \int{\sqrt{\lambda(p,n)} \over \sqrt{\lambda(p,D)}}
\nonumber \\
&\times&
 \rho_{\mu\nu}(D) \left[\rho_{pN}^\pi \cdot \Phi^{\mu\nu}(D,q)\right]
 {m^2d^3p' \over E'}~,
 \label{ND2}
 \end{eqnarray}
 where $\lambda(p_1, p_2) \equiv (p_1p_2)^2 - m_1^2 m_2^2 =
 \lambda(s_{12}, m_1^2, m_2^2) / 4$ is the flux factor; $p,n$ are
 the four-momenta of the proton-target and intra-deuteron nucleon,
 respectively;
 $\rho_{pN}^\pi \equiv \varepsilon_\pi {d\sigma / d^3p_\pi}$ is the
 relativistic invariant inclusive spectrum of pions produced by interacting
 the intra-deuteron nucleon with the proton-target. In the general case, this
 spectrum can be written as a three-variable function
 $\rho_{pN}^{\pi} = \rho(x_{\rm F}, \pi_\perp, s_{NN})$.
 The Feynman variable, $x_{\rm F}$, is defined as
 $x_{\rm F} = 2\pi_{||} / \sqrt{s_{NN}}$, where $\pi$ is the pion momentum
 in the center of mass of two interacting nucleons and $s_{NN} = (p + N)^2$.

 $\rho_{\mu\nu}(D)$ is the density matrix of the deuteron \cite{bon96}:
 \begin{eqnarray}
 &\rho_{\mu\nu}(D)& ~=~
         \left(\xi_{\mu M}(D)\right)^*\xi_{\nu M}(D)
 \label{dens_matr} \\
         &=&{1\over3}\left(-g_{\mu\nu}+{D_\mu D_\nu\over M^2}\right)+
         {1\over2}(W_\lambda)_{\mu\nu}s_D^\lambda
 \nonumber \\
 &-&\biggl[{1\over2}\biggl(
   (W_{\lambda_1})_{\mu\rho}(W_{\lambda_2})^{\rho}_{~\nu}+
  (W_{\lambda_2})_{\mu\rho}(W_{\lambda_1})^{\rho}_{~\nu}\biggr)
\nonumber \\
  &+&{2\over3}\left(-g_{\lambda_1\lambda_2} +
                 {D_{\lambda_1}D_{\lambda_2}\over M^2}\right)
  \left(-g_{\mu\nu} + {D_\mu D_\nu\over M^2}\right)\biggl]
  p_D^{\lambda_1\lambda_2}
\nonumber
 \end{eqnarray}
 with $(W_\lambda)_{\mu\nu}=i\varepsilon_{\mu\nu\gamma\lambda}D^\gamma/M$;
 $s_D$ the spin vector and $p_D$ the alignment tensor of the deuteron.

 The full symmetric tensor $\Phi_{\mu\nu}(D,q)$ in Eq.(\ref{ND2}) reads as
 \begin{eqnarray}
 \Phi_{\mu\nu}(D,q) &=&
 {1\over4}Tr\left[\bar\Psi_\mu\left({\widehat n+m\over m}\right)^2
 \Psi_\nu{\widehat{p'}-m\over m}\right]
\nonumber \\
 &=& -f_0(n^2)g_{\mu\nu} + f_1(n^2){q_\mu q_\nu\over m^2}~.
 \label{ND3}
 \end{eqnarray}
 Proving Eq.(\ref{ND3}), we introduce the modified vertex,
 $\Psi_\mu(D,q)$:
 \begin{eqnarray}
 \Psi_\mu(D,q) = {\Gamma_\mu(D,q)\over m^2-n^2-i0}=
 \varphi_1(n^2)\gamma_\mu + \varphi_2(n^2){n_\mu\over m}
\nonumber \\
  +{\widehat n-m\over m}
 \left(\varphi_3(n^2)\gamma_\mu + \varphi_4(n^2){n_\mu\over m}\right).
 \label{WFD}
 \end{eqnarray}

 Substituting Eq.(\ref{WFD}) into Eq.(\ref{ND3}) and calculating the
 trace, one can find explicit forms of the invariant functions $f_{0,1}$:
 \begin{eqnarray}
 f_0(n^2) &=&
 {M^2\over m^2}\left(\varphi_1-{m^2-n^2\over m^2}\varphi_3\right)\varphi_1
\nonumber \\
 &-& \left({m^2-n^2\over m^2}\right)^2
  \left(\varphi_1-\varphi_3\right)\varphi_3~;
\nonumber \\
 f_1(n^2) &=&
 -4\left[\varphi_1+\varphi_2-{m^2-n^2\over m^2}\left({\varphi_2\over2}+
 \varphi_3+\varphi_4\right)\right]
\nonumber \\
 &\times& \left(\varphi_1+\varphi_2\right)
 + {M^2\over m^2}\left(\varphi_2-{m^2-n^2\over m^2}\varphi_4\right)\varphi_2
\nonumber\\
 &-& \left({m^2-n^2\over m^2}\right)^2\left(\varphi_2+2\varphi_3+\varphi_4\right)\varphi_4~.
 \label{ND4}
 \end{eqnarray}
 The corresponding invariant scalar functions $\varphi_i(n^2)$ of the
 deuteron vertex with one on-shell nucleon can be computed in any reference
 frame. Let us note that in our case, when one particle is on-mass shell,
 only four partial amplitudes contribute to the process, namely in the
 $\rho$-spin classification,
 $U=^3{\cal S}^{++}_1, W=^3{\cal D}^{++}_1, V_s=^1{\cal P}^{-+}_1$ and
 $V_t=^3{\cal P}^{-+}_1$.  We can write $\varphi_i$ in the deuteron rest
 frame in order to relate them to the non-relativistic ${\cal S}$- and
 ${\cal D}$-waves of the deuteron. In this case, the invariant functions
 take the following forms:
\begin{eqnarray}
 {\cal N}\varphi_1 &=& U-{W\over\sqrt2}-
              \sqrt{3\over2}{m\over|\vec q|}V_t~;
\nonumber\\
 {\cal N}\varphi_2 &=& -{m\over(E_{\vec q}+m)}U-
             {m(2E_{\vec q}+m)\over|\vec q|^2}{W\over\sqrt2}+
             \sqrt{3\over2}{m\over|\vec q|}V_t~;
\nonumber\\
 {\cal N}\varphi_3 &=&
 -\sqrt{3\over2}{mE_{\vec q}\over|\vec q|(2E_{\vec q}-M)}V_t~;
\nonumber\\
 {\cal N}\varphi_4 &=&{m^2\over M(E_{\vec q}+m)}U-
             {m^2(E_{\vec q}+2m)\over M|\vec q|^2}{W\over\sqrt2}
\nonumber\\
                   &-&\sqrt3{m^2\over|\vec q|(2E_{\vec q}-M)}V_s~,
\label{BS}
\end{eqnarray}
 where all the vertex functions are determined in the deuteron rest frame and
 all the kinematic variables in Eqs.(\ref{BS}) have to be evaluated in this
 system; $E_{\vec q}=\sqrt{|\vec q|^2+m^2}$.
 The normalization factor ${\cal N}^{-1}=\pi\sqrt{2/M}$ is chosen
 according to non-relativistic normalization DWF:
 \footnote{
 Note, the Gross definition \cite{gro79} of DV
 $\widetilde \Psi_\mu(q)$ is related to the BS vertex
 $\Psi_\mu(q)$ (\ref{WFD}) as the follows:
 \begin{eqnarray}
 \widetilde \Psi_\mu(q) = \Psi_\mu(-q)~.
 \label{GROSS}
 \end{eqnarray}
 Comparing it with Eqs.(46) of Ref. \cite{gro79}, one can see that the Gross
 wave functions $\widetilde U(q), \widetilde W(q)$ and
 $\widetilde V_s(q), \widetilde V_t(q)$ are connected with our wave functions
 (\ref{BS}) as
 \begin{eqnarray}
 &&\widetilde U(q_0, |\vec q|) = U( - q_0, |\vec q|)~;~~
   \widetilde W(q_0, |\vec q|) = W( - q_0, |\vec q|)~;~~
 \nonumber \\
 &&\widetilde V_s(q_0, |\vec q|) = - V_s( - q_0, |\vec q|)~;~~
   \widetilde V_t(p_0, |\vec p|) = - V_t( - q_0, |\vec p|)~,
 \label{GROSSCONN}
 \end{eqnarray}
 where  $q_0 = M/2 - E_{\vec q}$.
 }
 $$\int\limits_0^\infty |\vec q|^2d|\vec q|
   \left(U^2(|\vec q|) + W^2(|\vec q|)\right) = 1$$
 The relativistic invariant functions $f_{0,1}(|\vec q|)$ (\ref{ND4})
 can be rewritten in terms of these spin-orbit momentum wave functions as
\begin{eqnarray}
 f_0(|\vec q|) &=&
 {\cal N}^{-2}{M^2\over m^2}\biggl[\left(U-{W\over\sqrt2}\right)^2
\nonumber \\
 &+&\sqrt6
  {|\vec q|\over m}\left(U-{W\over\sqrt2}\right)V_t-{3\over2}V_t^2\biggl]~;
\label{ND5} \\
 f_1(|\vec q|) &=&
 {\cal N}^{-2}{3M^2\over 2|\vec q|^2}\biggl[2\sqrt2UW+W^2+V_t^2-2V_s^2
\nonumber \\
 &-& {4\over\sqrt3}{|\vec q|\over m}\left(\left(U-{W\over\sqrt2}\right)
 {V_t\over\sqrt2}+\left(U+\sqrt2W\right)V_s\right)\biggl].
\nonumber
 \end{eqnarray}
 Then, all the observables can be computed in terms of positive- and
 negative-energy wave functions $U,W$ and $V_s,V_t$, respectively.
 The contribution of the positive-energy waves $U,W$ to the observables
 results in the non-relativistic limit. The parts containing
 the negative-energy waves $V_s,V_t$ have a pure relativistic
 origin, and consequently they manifest genuine relativistic correction
 effects.

 Using an explicit form of the density matrix (\ref{dens_matr}), one can
 write
 \begin{eqnarray}
 \Phi\equiv\rho_{\mu\nu}\Phi^{\mu\nu} = \Phi^{(u)} +
 \Phi^{(v)}_\lambda s_D^\lambda +
 \Phi^{(t)}_{\lambda_1\lambda_2}p_D^{\lambda_1\lambda_2}~.
 \label{ND6}
 \end{eqnarray}
 The superscripts $(u,v,t)$ denote unpolarized, vector polarized and
 tensor polarized cases, respectively:
\begin{eqnarray}
 &&\Phi^{(u)}(\vec q)=f_0+{1\over3}{|\vec q|^2\over m^2}f_1~;
\nonumber \\
 &&\Phi^{(v)}_\lambda(\vec q)=0~;
\label{ND7} \\
 &&\Phi^{(t)}_{\lambda_1\lambda_2}(\vec q)=
 \biggl[{1\over3}{|\vec q|^2\over m^2}
 \left(-g_{\lambda_1\lambda_2}+{D_{\lambda_1}D_{\lambda_2}\over M^2}\right)
\nonumber\\
 &&-\left(-g_{\lambda_1\mu}+{D_{\lambda_1}D_\mu\over M^2}\right)
 \left(-g_{\lambda_2\nu}+{D_{\lambda_2}D_\nu\over M^2}\right)
 {q^\mu q^\nu\over m^2}\biggl] f_1.                          \nonumber
 \end{eqnarray}
 Let us consider now the case when the deuteron has a tensor polarization.
 If the initial deuteron is only aligned due to the $p_D^{zz}$ component,
 then the inclusive spectrum of the reaction
 $\vec D + p \to \pi + X$ (\ref{ND2}) can be written in the form:
 \begin{eqnarray}
 \rho_{pD}^\pi \left(p_D^{ZZ}\right)=
 \rho_{pD}^\pi \left[1 + {\rm A}_{ZZ} ~ p_D^{ZZ}\right]~,
 \label{ND8}
 \end{eqnarray}
 where $\rho_{pD}^\pi$ is the inclusive spectrum for the case of unpolarized
 deuterons and ${\rm A}_{ZZ} \equiv \sqrt2T_{20}$
 $(-\sqrt2 \leq T_{20} \leq 1/\sqrt2)$ is the tensor analyzing power.
 One can write:
\begin{eqnarray}
 \rho_{pD}^\pi &=&
 {1\over(2\pi)^3} \int {\sqrt{\lambda(p,n)}\over\sqrt{\lambda(p,D)}}\left[
 \rho_{pN}^\pi\cdot\Phi^{(u)}(|\vec q|)\right]{m^2d^3q\over E_{\vec q}};~
\label{ND9} \\
 \rho_{pD}^\pi &\cdot& {\rm A}_{ZZ} =
 - {1\over(2\pi)^3} \int {\sqrt{\lambda(p,n)}\over\sqrt{\lambda(p,D)}}
\nonumber \\
 &\times& \left[\rho_{pN}^\pi\cdot\Phi^{(t)}(|\vec q|)\right]
 \left({3\cos^2\vartheta_{\vec q}-1\over2}\right){m^2d^3q\over E_{\vec q}}~,
\label{ND10}
\end{eqnarray}
 where
\begin{eqnarray}
 &&\Phi^{(u)}(|\vec q|) = {\cal N}^{-2} {M^2\over m^2}
 \biggl[ U^2+W^2 - V_t^2 - V_s^2
\nonumber \\
 &&~+ {2\over\sqrt3}{|\vec q|\over m} \left(
 \left(\sqrt2V_t-V_s\right)U - \left(V_t+\sqrt2V_s\right)W\right)\biggl];~~
\label{ND11} \\
 &&\Phi^{(t)}(|\vec q|) = {\cal N}^{-2} {M^2\over m^2}
 \biggl[ 2\sqrt2UW + W^2 + V_t^2 - 2V_s^2
\nonumber \\
 &&~- {4\over\sqrt3}{|\vec q|\over m}\left(\left(U-{W\over\sqrt2}\right)
 {V_t\over\sqrt2} + \left(U+\sqrt2W\right)V_s\right)\biggl].~~~
\label{ND12}
\end{eqnarray}

 It is intuitively clear that the nucleons in the deuteron are mainly
 in the states with angular momenta $L=0,2$ (see also a numerical analysis
 of the solutions of the $BS$ equation in terms of amplitudes within the
 $\rho$-spin basis \cite{kap96}), the probability of states with $L=1$
 $(V_{s,t})$ in Eqs.(\ref{ND11},\ref{ND12}) is much smaller than the
 probability for the $U,W$ configurations. Moreover, it can be shown that
 the $U$ and $W$ waves directly correspond to the non-relativistic ${\cal S}$
 and ${\cal D}$ ones. Therefore, Eqs.(\ref{ND11},\ref{ND12}) with only $U,W$
 waves can be identified as main contributions to the corresponding
 observables, and they can be compared with their non-relativistic analogies.
 The other terms possessing contributions from ${\cal P}$-waves are
 proportional to $\vec q/m$ (the diagonal terms in $V_{s,t}$ are negligible).
 Due to their pure relativistic origin, one can refer to them as relativistic
 corrections.

 Let us consider a minimal relativization scheme which describes rather well
 the differential cross section for such a process as deuteron break-up
 $A(D,p)X$. The minimal relativization procedure \cite{bro73,fra81}
 comprises the following (i) replacing the argument of the non-relativistic
 wave functions by the light-cone variable
 $\vec k=(\vec k_\perp,\vec k_{||})$
\begin{equation}
 \vec k^2={m^2+\vec k_\perp^2\over4x(1-x)}-m^2;~
 k_{||}=\sqrt{m^2+\vec k_\perp^2\over x(1-x)}\left({1\over2}-x\right).
\label{lcv}
\end{equation}
 where
 $x = (E_{\vec q}+|\vec q|\cos\vartheta_{\vec q})/M
    = (\varepsilon'-p'_{||})/M;~|\vec k_\perp| = p'_\perp$ in the deuteron
 rest frame, and (ii) multiplying the
 wave functions by the factor $\sim 1/(1-x)$. It results in a shift
 of the argument towards larger values, and the wave function itself
 decreases more rapidly. This effect of suppressing the wave function is
 compensated by the kinematic factor $1/(1-x)$.

 In the BS approach the relativistic effects are of a dynamic nature
 \cite{kap95} and not reduced to a simple shift in arguments. In addition
 to ${\cal S}$ and ${\cal D}$ waves, it contains negative energy components,
 i.e., ${\cal P}$ waves.
 One can see that they play a more important role in the polarization case
 and lead to an improvement of description of the data.

 \section{Results and discussion}
 \label{sec:res}

 Below, the calculated results of the inclusive relativistic
 invariant pion spectrum and the tensor analyzing power in the
 fragmentation process $D p \to\pi X$ are presented and compared with the
 available experimental data \cite{bal85,afa98}. These experimental data
 are presented as a functions of the so-called cumulative scaling
 variable $x_{\cal C}$ (``cumulative number'') \cite{sta79}. For our
 reaction, this variable is defined as follows:
\begin{eqnarray}
 x_{\cal C} &=& 2 {(p\pi) - \mu^2/2 \over (Dp) - Mm - (D\pi)}
\nonumber \\
            &=& 2 {t - m^2 \over (t - m^2) + (M + m)^2 - s_X} \leq 2~.
\label{xC}
\end{eqnarray}
 In the rest frame of the deuteron $D = (M, \vec 0)$, it can be rewritten
 in the form:
\begin{equation}
 x_{\cal C} = 2{EE_\pi - pp_\pi \cos\vartheta_\pi - \mu^2/2 \over
              M(E - E_\pi - m)}
          \to 2 {E \over T_p}{\alpha \over 1 - E_\pi/T_p}~,
\label{xClab}
\end{equation}
 where $\alpha = (E_\pi - p_\pi \cos\vartheta_\pi)/M$ is a light-cone
 variable.
 The value of $x_{\cal C}$ corresponds to a minimum mass (in nucleon mass
 units) of part of the projectile nucleus (deuteron) involved in the
 reaction. Values of $x_{\cal C}$ larger than $1$ correspond to cumulative
 pions.

 Deuteron (polarized and unpolarized) fragmentation into proton
 $D + A \to p(0^o) + X$ is one of the more intensively studied
 reaction with hadronic probe. The reason of this study is, first, a big
 cross section and, second, rather a simple relation of the inclusive
 spectrum and polarization observables to the ${\cal S}$- and
 ${\cal D}$-waves of the DWF obtained within IA. For example, the tensor
 analyzing power $T_{20}$ within IA can be written in the following simple
 form:
\begin{eqnarray}
 T_{20} = -{1 \over \sqrt{2}}~{2\sqrt{2}~U~W~+~W^2 \over U^2~+~W^2}
\label{t_20dp}
\end{eqnarray}
 This relation does not depend on the amplitude of elementary reaction
 $p n \to p X$ which is taken in IA as an off-mass shell. As
 shown in \cite{lyk93}, both the differential cross section and $T_{20}$ for
 fragmentation $D + p \to p(0^o) + X$ can be described within the IA
 at $k \leq 0.2$~GeV/c only. At larger momenta $k$, secondary
 interactions, in particular the triangle graphs with a virtual pion,
 have to be taken into account in order to describe these observables
 least satisfactorily.

 However, as shown in \cite{kon75} for the pion production
 $D + p \to \pi(0^o) + X$ at the cumulative over region the rescattering
 mechanism is kinematically suppressed.
 Therefore,  one can use the IA only by the theoretical calculus of the
 differential cross section and tensor analyzing power $T_{20}$ for this
 reaction.
 The calculated results of the invariant spectrum of pions produced by
 the $D + p \to \pi(0^o) + X$ reaction are presented in
 figs.~(\ref{fig:fig1}-\ref{fig:fig2}). The vertex $N N \to \pi Y$ is
 taken as an on-mass shell and corresponding differential cross section
 proposed in \cite{bar84} was used.

 A large sensitivity of the inclusive spectrum to this vertex and a small
 one to the type of the nonrelativistic DWF can be seen from
 fig.~\ref{fig:fig1}. The fig.~\ref{fig:fig2} shows that the inclusion of
 the ${\cal P}$-waves contribution to the DWF within the Bethe-Salpeter
 or Gross approaches results in a better (but not satisfactory) description
 of the experimental data over the cumulative region.
 From fig.~\ref{fig:fig3} one can see the effects of calculation using the
 minimal relativization scheme \cite{fra81}.

 The calculated results of $T_{20}$ for the reaction of polarized deuteron
 fragmentation into cumulative pions are shown in
 figs.~(\ref{fig:fig4}-\ref{fig:fig6}). From these figures one can see a
 small sensitivity of $T_{20}$ to the vertex corresponding to the
 $NN \to \pi Y$ process. It is also seen that $T_{20}$ is more sensitive
 to the DWF form than the invariant spectrum. The experimental data on
 $T_{20}$ are not described by any DWF used in this paper.
 Note that there may be an alternative approach to study the deuteron
 structure at small distances which assumes a possible existence of
 non-nucleon or quark degrees of freedom \cite{lyk93,glo93,kob93} in
 the deuteron and nucleus. For example, according to \cite{fra81},
 large momenta of nucleons are due to few nucleon correlations
 in the nucleus. Then the deuteron structure can be described by assuming
 the quark degrees of freedom \cite{luk79,bur84}. On the other hand,
 the shape of a high momentum tail of the nucleon distribution in
 the deuteron can be constructed from the basis of its true Regge asymptotic
 \cite{efr88}, and the corresponding parameters can be found from a
 good description of the inclusive proton spectrum in deuteron fragmentation
 $D p\to p X$ \cite{efr88,lyk93}. According to \cite{efr88,lyk93}, one can
 write the following form for $\widetilde\Phi^{(u)}(|\vec q|)$
 (see Eq.(\ref{ND11})):
 \begin{equation}
 \Phi^{(u)}(|\bm q|) = {E_{\bm k}/E_{\bm q} \over 2(1 - {\rm x})}
 \widetilde\Phi^{(u)}(|\bm k|)~.
 \label{Phi:6q-(q->k)}
 \end{equation}
where
\begin{eqnarray}
 \widetilde\Phi^{(u)}(|\bm k|) &=& {\cal N}^{-1} {M_D^2 \over m^2}
 \biggl[(1 - \alpha_{2(3q)}) \cdot \Bigl(U^2(|\bm k|) + W^2(|\bm k|)\Bigl)
\nonumber\\
 &+& \alpha_{2(3q)} {8\pi{\rm x}(1 - {\rm x}) \over E_{\bm k}}
 \cdot G_{2(3q)}({\rm x}, \bm k_\bot) \biggl].
\label{Phi:6q}
\end{eqnarray}
 The parameter $\alpha_{2(3q)}$ is the probability for a non-nucleon
 component in the deuteron which is a state of two colorless $(3q)$
 systems.
\begin{equation}
  G_{2(3q)}(x, \vec k_\bot) = {b^2 \over 2\pi} 
  {\Gamma(A + B + 2) \over \Gamma(A + 1)\Gamma(B + 1)}
  x^A (1 - x)^B ~ {\mbox{e}}^{-b k_\bot}.
\label{6q}
\end{equation}
 Fig.~\ref{fig:fig7} presents the invariant pion spectrum calculated within
 the relativistic impulse approximation including the non-nucleon component
 in the DWF \cite{lyk93,efr88}; its probability $\alpha_{2(3q)}$ is
 $0.2 \div 0.4$ (dot-dashed and dashed curves, respectively). One can see
 a good description of the experimental data \cite{bal85} at all
 $x_{\cal C}$.  
 The analogous results including non-nucleon degrees of freedom, 
 according to \cite{efr88,lyk93}, can be obtained for the tensor 
 analyzing power $T_{20}$. Actually, in \cite{efr88} a form of 
 $\widetilde\Phi^{(u)}(|\bm k|)$ has been constructed only. However,
 to calculate $T_{20}$ it is not enough, the corresponding orbital waves
 have to be known. Let us assume, the non-nucleon degrees of freedom
 result in a main contribution to the ${\cal S}$- and ${\cal D}$-waves of
 the deuteron wave function. Constructing new forms of these waves
 by including the non-nucleon degrees of freedom we have to require
 that the square of the new DWF has to be equal to the one determined
 by the Eq.(\ref{Phi:6q}).
 Introducing a mixing parameter $\alpha=\pi a/4$ one can find the forms 
 of new ${\cal S}$- and ${\cal D}$-waves as the following:
 \begin{eqnarray}
  \widetilde{U}(|\bm k|) &=&
   \sqrt{1-\alpha_{2(3q)}}U(|\bm k|) + \cos(\alpha) \Delta(|\bm k|)~;
 \label{def:Unew} \\
  \widetilde{W}(|\bm k|) &=&
   \sqrt{1-\alpha_{2(3q)}}W(|\bm k|) + \sin(\alpha) \Delta(|\bm k|)~,
 \label{def:Wnew}
 \end{eqnarray}
 where the function $\Delta$
 has been obtained from the equation:
 \begin{equation}
  \widetilde\Phi^{(u)}(|\bm k|) ~=~
  {\cal N}^{-1}{M_D^2 \over m^2}
   \biggl[\widetilde{U}^2(|\bm k|) + \widetilde{W}^2(|\bm k|)\biggl]~.
 \label{rel:UW-Phi}
 \end{equation}

 Fig.~\ref{fig:fig8} presents the analyzing power $T_{20}$ including
 calculated by using the functions $\widetilde{U}, \widetilde{W}$
 including the non-nucleon components in the DWF, according to
 \cite{lyk93,efr88}.
 It is shown from fig.~\ref{fig:fig8} the inclusion of non-nucleon
 components in the DWF improves the description of the experimental data 
 about $T_{20}$ at $x_C > 1$. The value of the parameter entering the
 Eqs.(\ref{def:Unew},\ref{def:Wnew}) $a=2.3$ results in an optimal
 description of this observable.

 \section{Summary and outlook}
 \label{sec:sum}

 The main goal of this paper is to study the reaction of deuteron
 fragmentation into pions within the framework of the nucleon model of
 deuteron and find a role of the non-nucleon degrees of freedom in a
 deuteron in this process. Main results can be summarized as follows.

\begin{figure*}
\centerline{\epsfxsize=17cm \epsffile{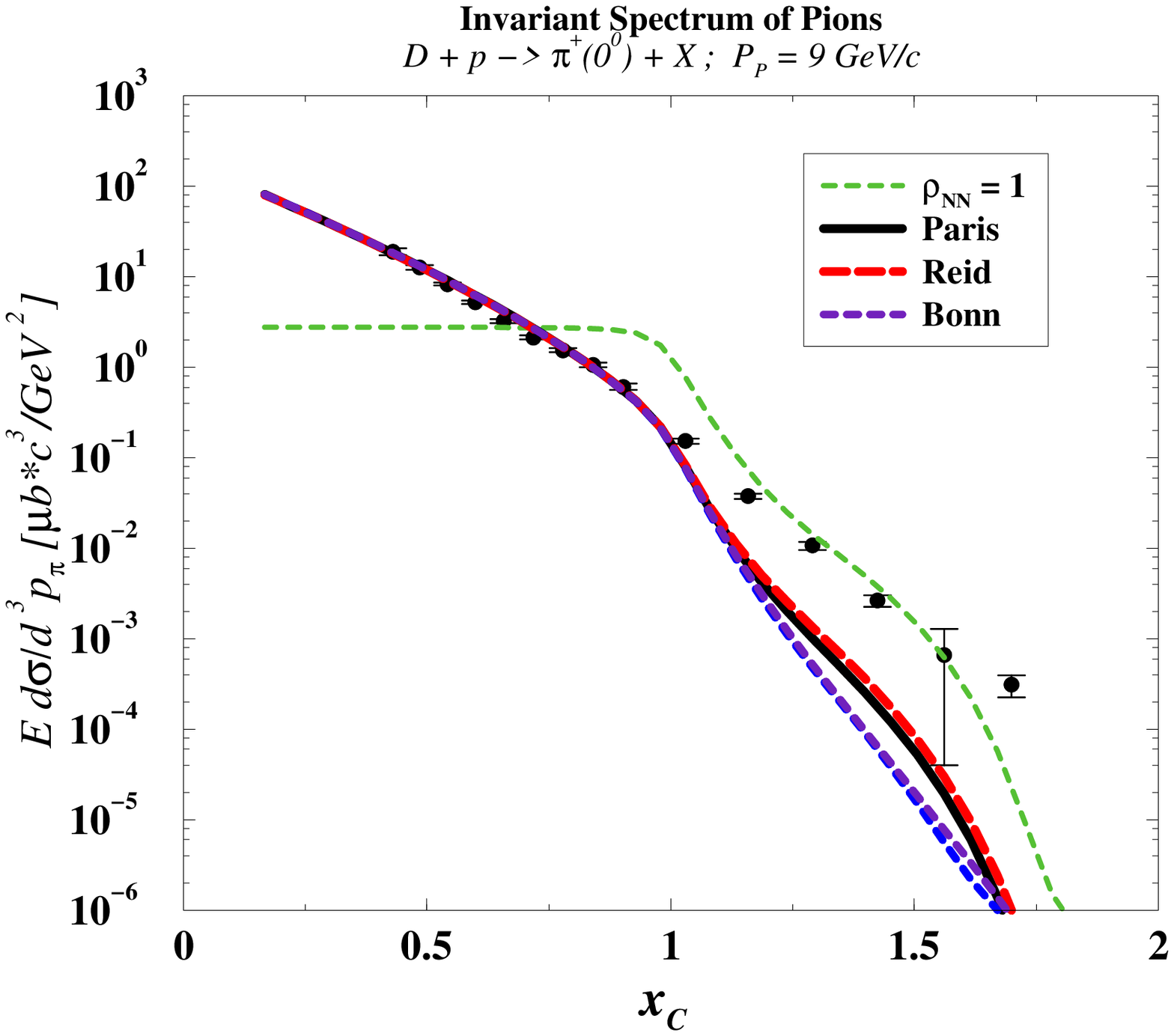}}
\vspace*{4cm}       
\caption{
 The invariant spectrum of the backward pions in the deuteron
 fragmentation reaction calculated in the relativistic impulse approximation
 using various types of the nonrelativistic DWF. The calculated results
 are compared with the experimental data from \protect\cite{bal85} for a
 the projectile proton momentum of $P_p = 9$~GeV/c. The thin dashed curve
 corresponds to the calculus by neglecting the dependence of the elementary
 vertex $N N \to \pi Y$ on the relativistic invariant variables.
}
\label{fig:fig1}       
\end{figure*}

\begin{figure*}
\centerline{\epsfxsize=17cm \epsffile{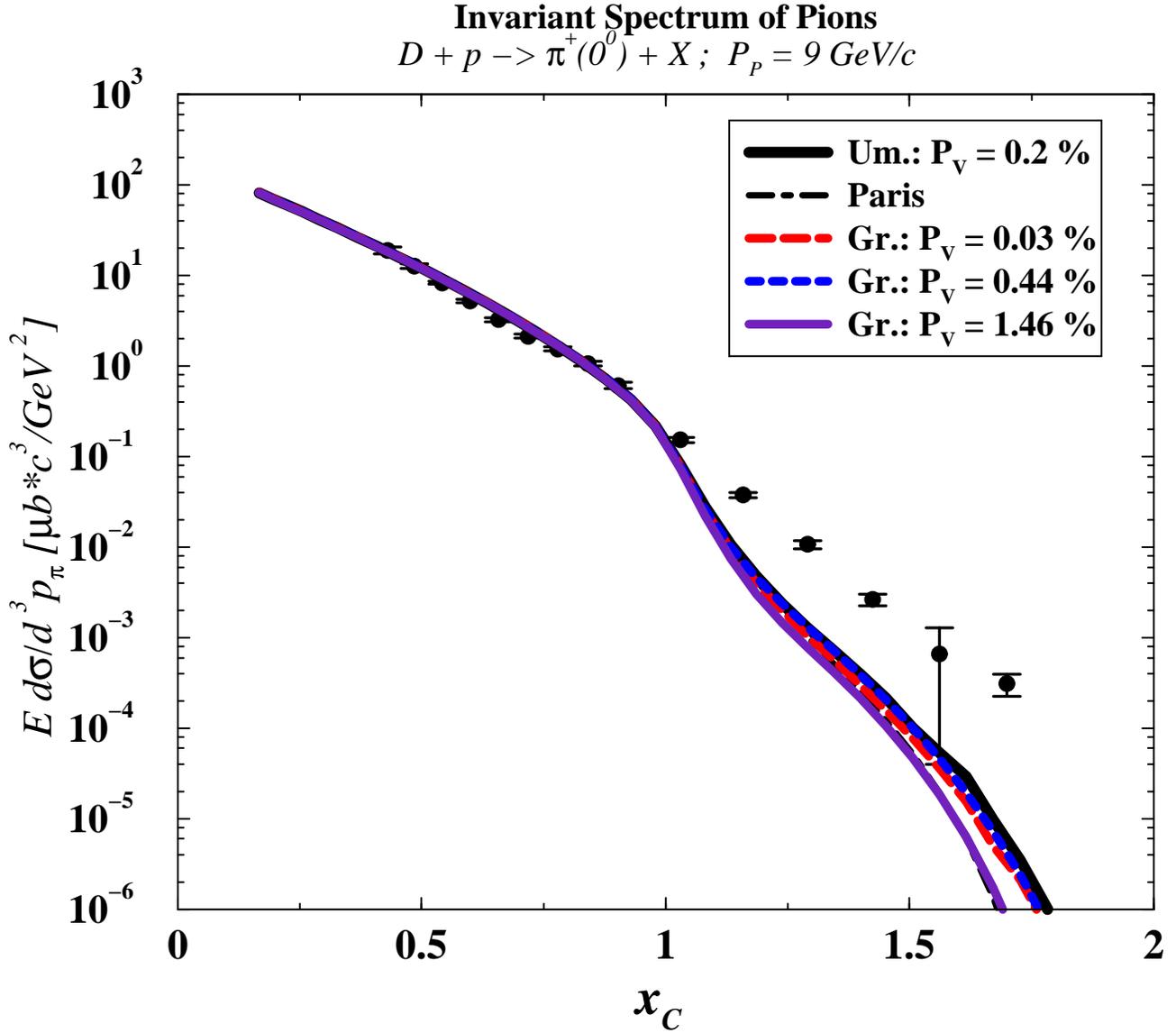}}
\vspace*{4cm}       
\caption{
 The invariant spectrum calculated using two forms of the
 relativistic DWF's, \protect\cite{umn96} and \protect\cite{gro79}.
 The experimental data  are taken from \protect\cite{bal85}.
 The thin solid line corresponds to the DWF
 \protect\cite{umn96}, where the total probability of small components:
 $P_{V} = \int_0^\infty p^2dp \cdot[V_t^2 + V_s^2]\simeq 0.2~\%$.
 The long-dashed, dashed and thick solid lines represent the calculations
 with the Gross DWF using the mixing parameter $\lambda = 0.0, 0.4$ and
 $1.0$, respectively \protect\cite{gro92}. This corresponds
 probabilities $P_{v} = 0.03~\%,~0.44~\%$ and $1.46~\%$
 to be obtained of the small component. The dot-dashed line
 corresponds to the nonrelativistic Paris DWF.
}
\label{fig:fig2}       
\end{figure*}

\begin{figure*}
\centerline{\epsfxsize=17cm \epsffile{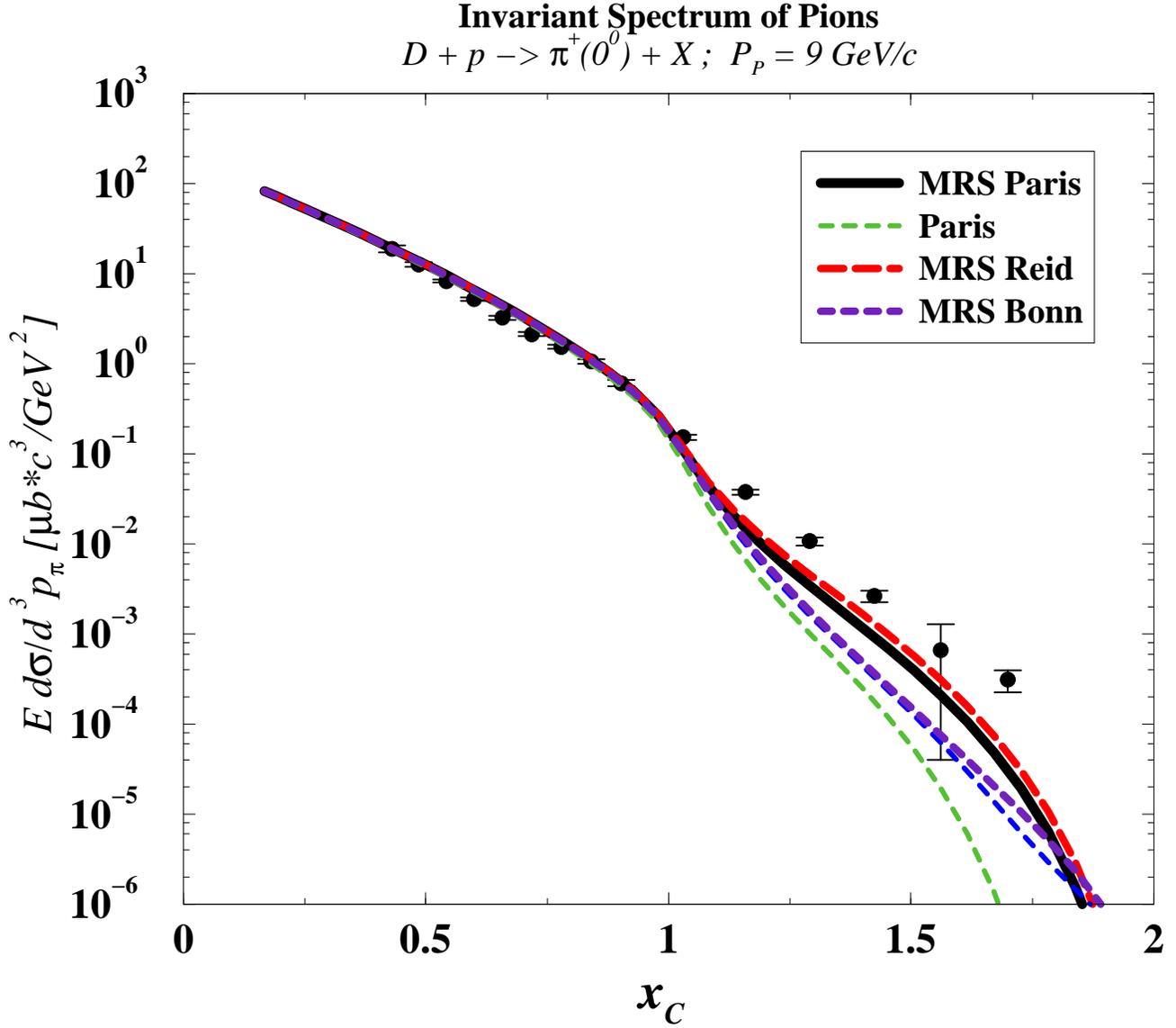}}
\vspace*{4cm}       
\caption{
 The invariant pion spectrum calculated using the
 nonrelativistic DWF's obtained by the minimal relativization scheme (MRS)
 \protect\cite{fra81,bro73}. The solid, dashed and long-dashed lines
 correspond to various DWF forms: Paris, RSC and Bonn.
 The experimental data are taken from \protect\cite{bal85}.
}
\label{fig:fig3}       
\end{figure*}

\begin{figure*}
\centerline{\epsfxsize=17cm \epsffile{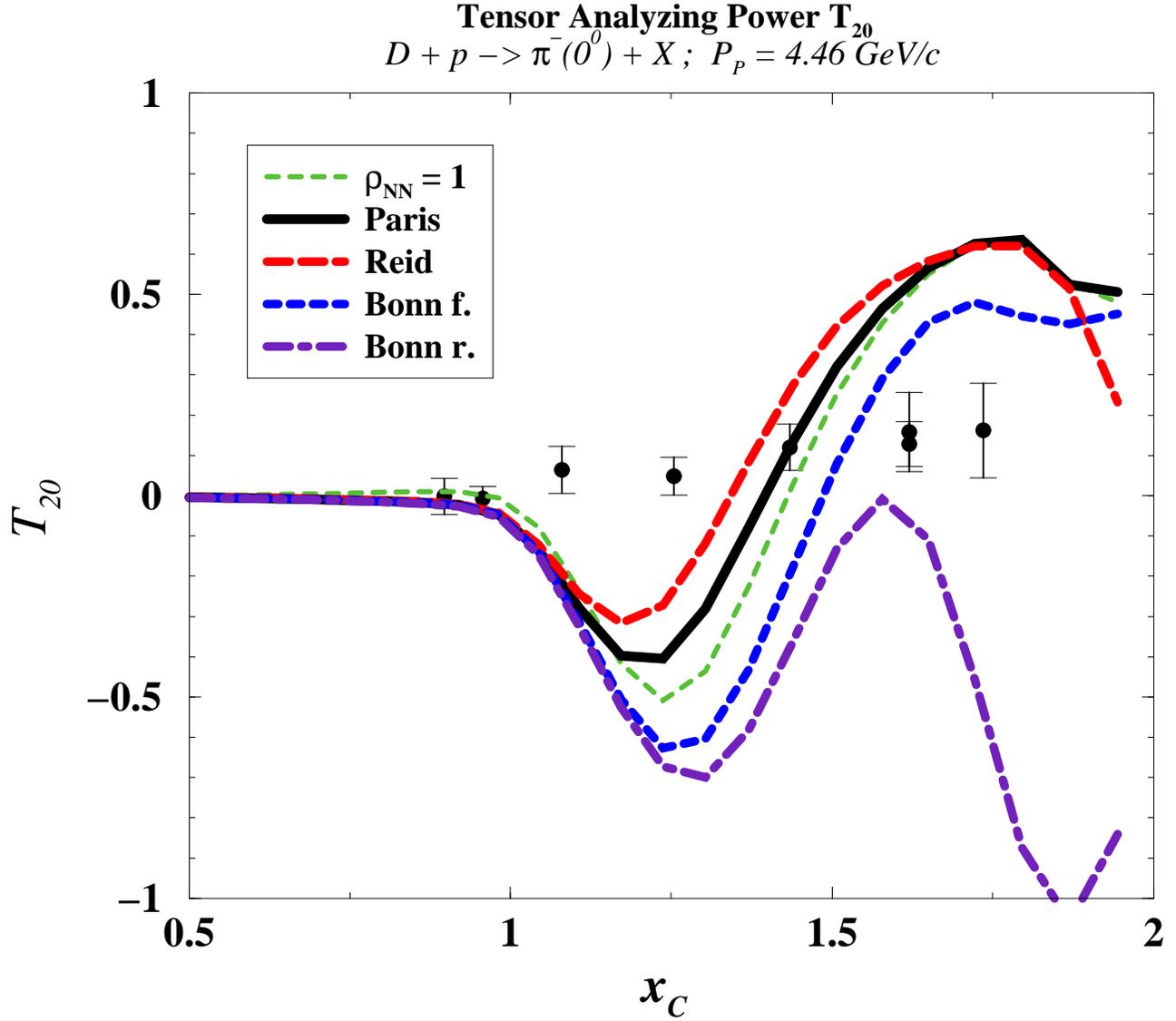}}
\vspace*{4cm}       
\caption{
 The tensor analyzing power $T_{20}$ of deuterons.
 The calculated results are  compared with the experimental data from
 \protect\cite{afa98} at a projectile proton momentum $P_p = 4.46$~GeV/c.
 The thin dashed curve corresponds to the calculus by neglecting the
 internal structure of the elementary vertex $NN \to \pi Y$. The solid,
 long-dashed, dashed and dot-dashed lines correspond to the calculus
 using various types of the nonrelativistic DWF: Paris, RSC
 and two Bonn types -- relativistic Bonn DWF and full Bonn DWF,
 respectively.
}
\label{fig:fig4}       
\end{figure*}

\begin{figure*}
\centerline{\epsfxsize=17cm \epsffile{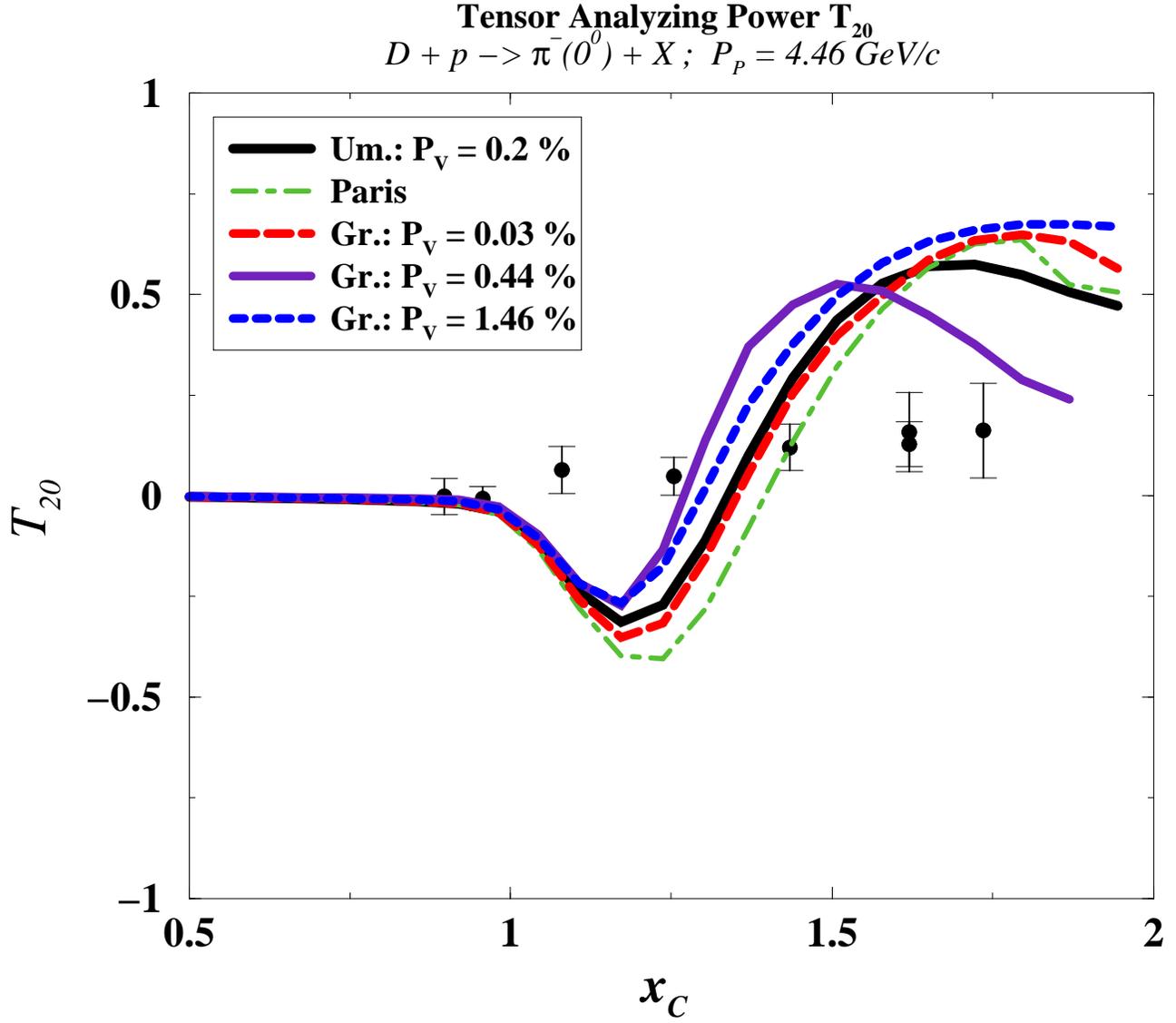}}
\vspace*{4cm}       
\caption{
 $T_{20}$ calculated using two forms of the relativistic DWF's,
 \protect\cite{umn96} and \protect\cite{gro79}.
 Notation as in fig.~\ref{fig:fig3} .
 The experimental data are taken from \protect\cite{afa98}.
}
\label{fig:fig5}       
\end{figure*}

\begin{figure*}
\centerline{\epsfxsize=17cm \epsffile{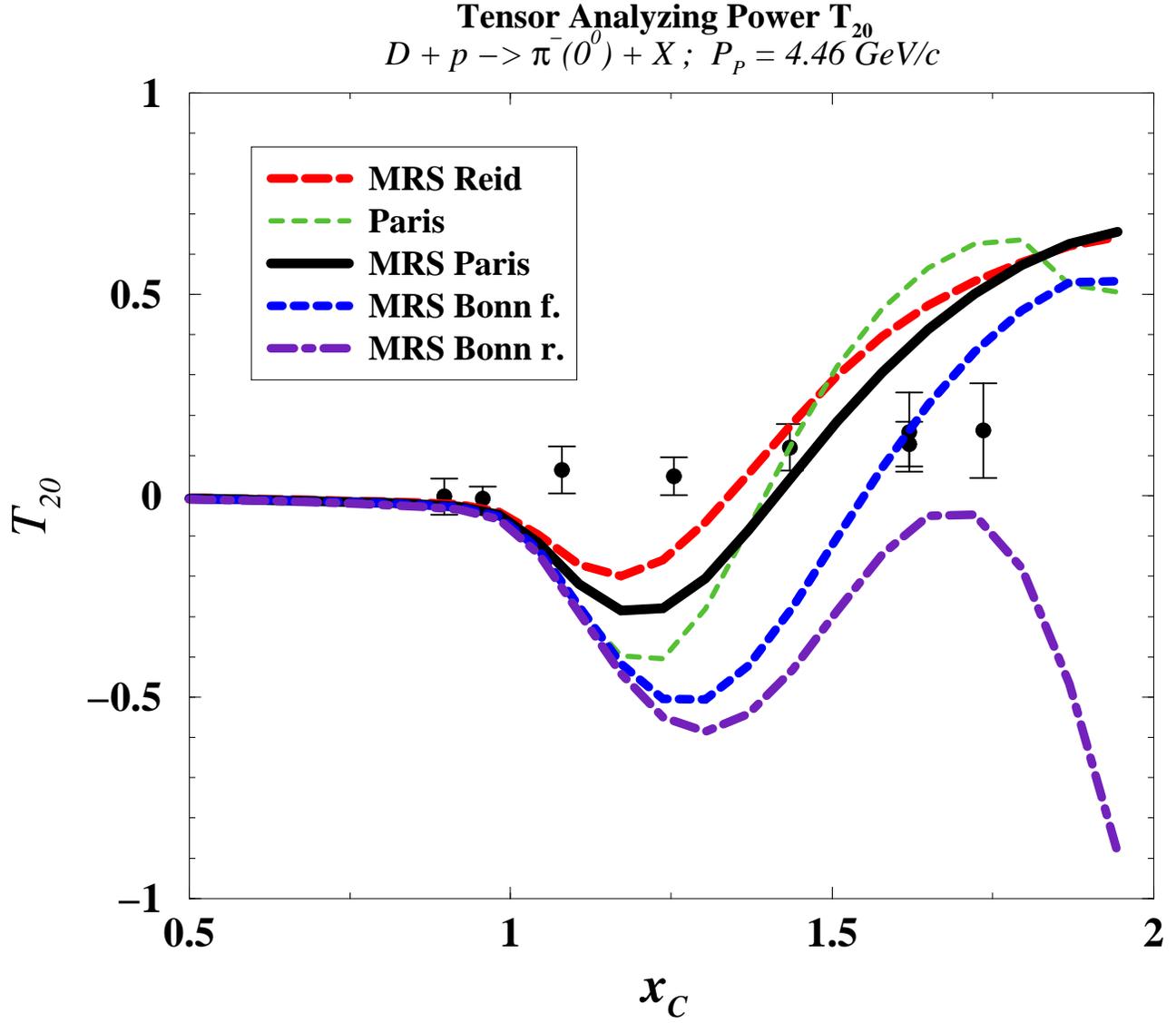}}
\vspace*{4cm}       
\caption{
 $T_{20}$ calculated with the nonrelativistic DWF's
 using the minimal relativization scheme (MRS) \protect\cite{fra81,bro73}.
 The solid, thick dashed, thin dashed and long-dashed lines correspond to
 the various DWF forms: Paris, the RSC and two Bonn DWF's --
 full and relativistic. The experimental data are taken
 from \protect\cite{afa98}.
}
\label{fig:fig6}       
\end{figure*}

\begin{figure*}
\centerline{\epsfxsize=17cm \epsffile{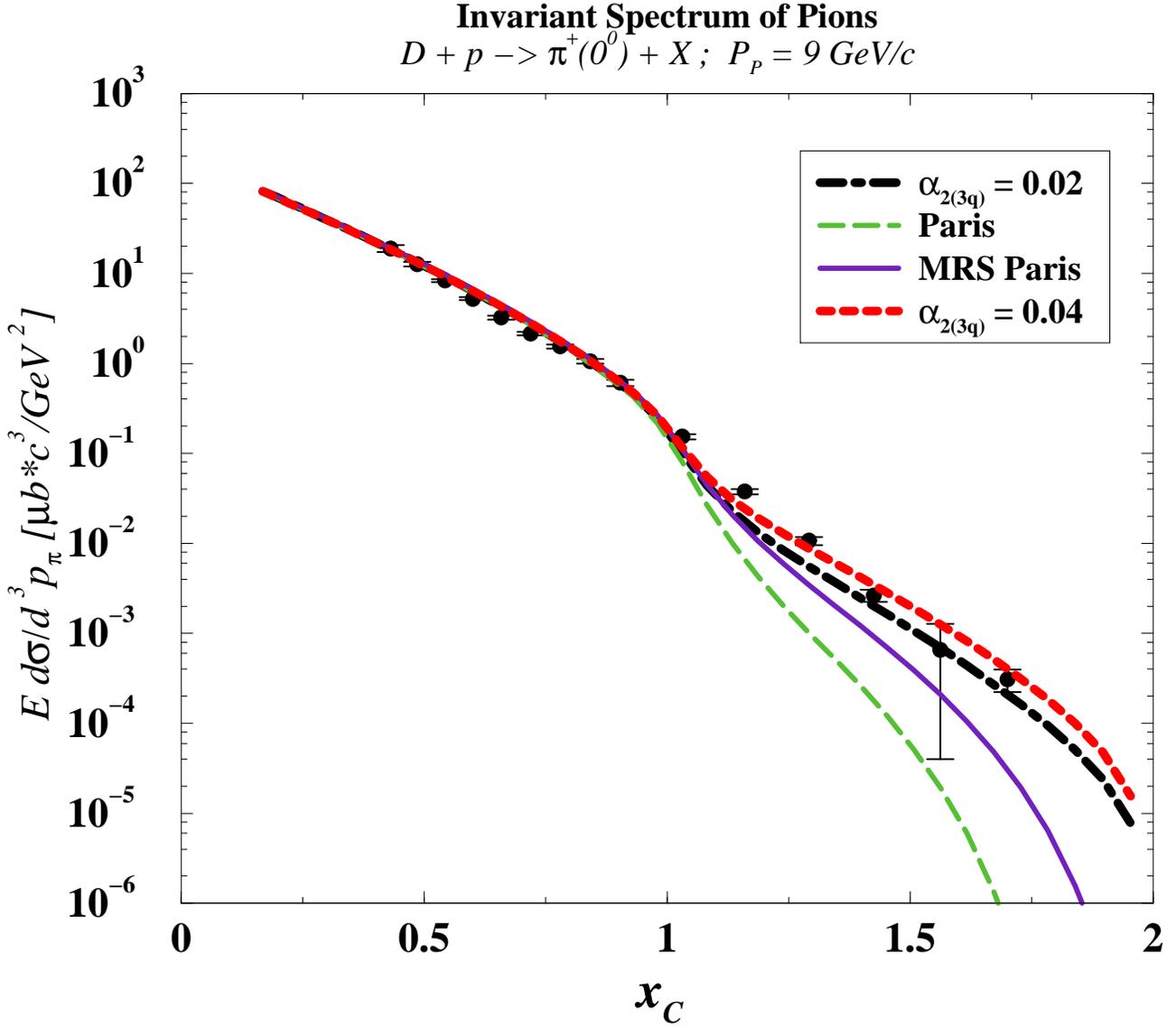}}
\vspace*{4cm}       
\caption{
 The invariant pion spectrum calculated within the relativistic
 impulse approximation with including of the non-nucleon component in the DWF
 \protect\cite{efr88,lyk93}; its probability $\alpha_{2(3q)}$ is
 $0.02\div0.04$ (dot-dashed and dashed curves, respectively).
 One can have a good description of the experimental data
 \protect\cite{bal85} for all $x_{\cal C}$.
}
\label{fig:fig7}       
\end{figure*}

\begin{figure*}
\centerline{\epsfxsize=17cm \epsffile{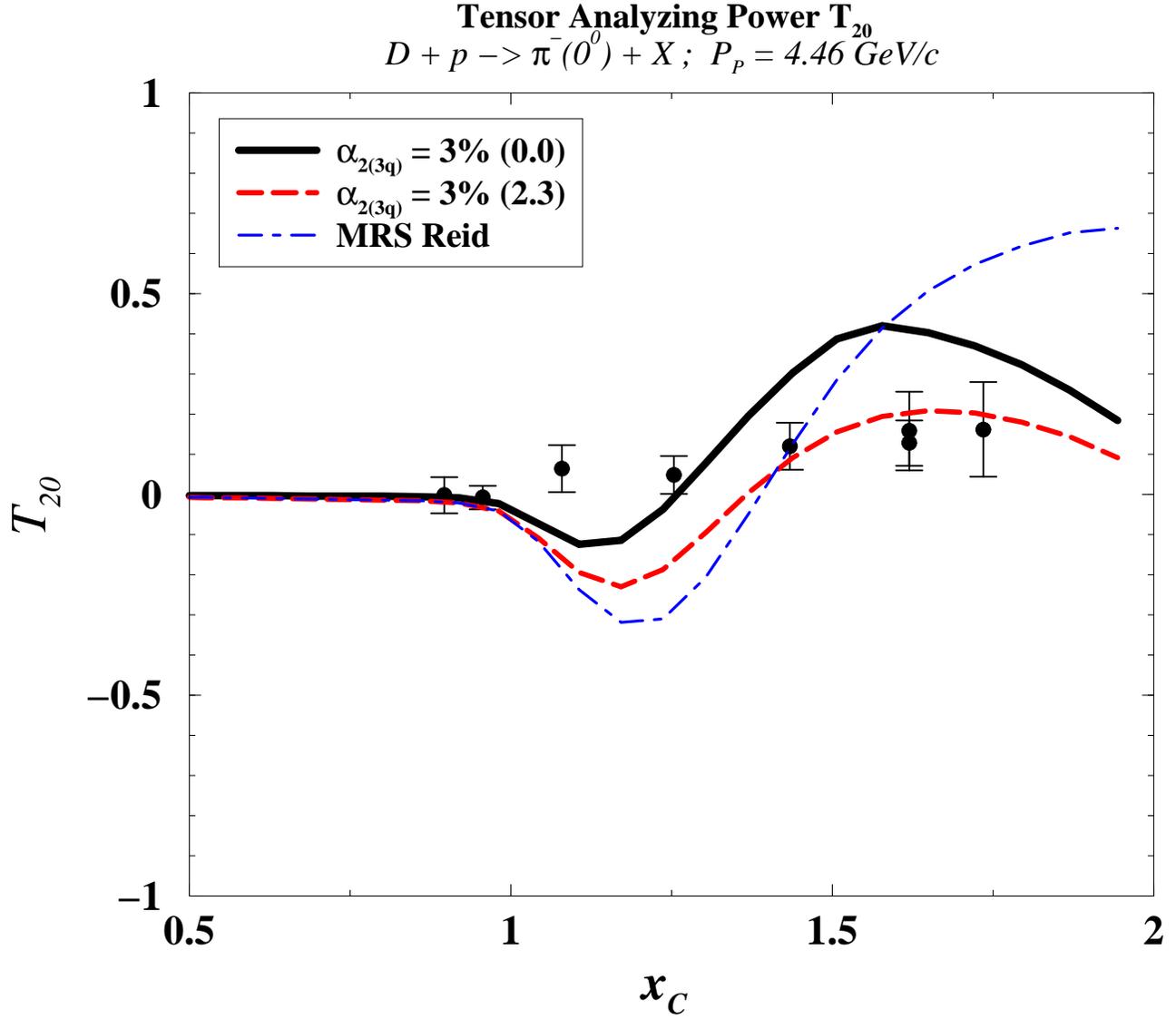}}
\vspace*{4cm}       
\caption{
 The tensor analyzing power $T_{20}$ calculated within the relativistic
 impulse approximation including of the non-nucleon component in the DWF
 \protect\cite{efr88,lyk93} with probability $\alpha_{2(3q)} = 0.03$.
 The solid and dashed lines represent the calculations using the mixing
 parameter $a = 0.0$ and $2.3$, respectively,
 Eqs.~(\ref{def:Unew},\ref{def:Wnew}).
}
\label{fig:fig8}       
\end{figure*}

 \begin{enumerate}
 \item It is quite incorrect to use the nonrelativistic deuteron wave
 function
 for the analysis of $D - N$ fragmentation into hadrons, in particular pions.
 Relativistic effects are sizable, especially in the kinematic region
 corresponding to short intra-deuteron distances or large $x$. It is seen
 from the behavior of the inclusive pion spectrum and particularly the tensor
 analyzing power $T_{20}$ at large $x$.

 \item At the present time, the state of theory is such, that the unique
 procedure to include relativistic effects in the deuteron has not been
 found yet. An extreme sensitivity to different methods of the
 relativization deuteron wave function is found for $T_{20}$ at $x \geq 1$.

 \item A large sensitivity of the inclusive spectrum of pions to the vertex
 of the $NN \to \pi X$ process can be seen from fig.~\ref{fig:fig1}.
 In contrast to this the small sensitivity of $T_{20}$ to this vertex is
 found, as seen from fig.~\ref{fig:fig4}. This polarization observable is
 very sensitive to the DWF form (figs.~(\ref{fig:fig4}-\ref{fig:fig6})).

 \item Very interesting experimental data on $T_{20}$ showing
 approximately zero values at $x_{\cal C}\geq 1$ are not reproduced by
 a theoretical calculus using even different kinds of the relativistic
 DWF. This may indicate a possible existence of non-nucleon degrees
 of freedom or basically new mechanism of pion production in the kinematic
 region forbidden for free $N - N$ scattering.

 \item For the deuterons fragmentation into protons emitted forward,
 the tensor analyzing power $T_{20}$ is not described by the standard
 nuclear physics using the nucleon degrees of freedom at
 $x_{\cal C} \geq 1.7$ \cite{lyk93}.
 On the contrary, $T_{20}$ for the fragmentation $D p \to \pi X$
 can not be described within the same assumptions over all region
 $x_{\cal C} \geq 1$.
 The inclusion of the non-nucleon degrees of freedom within the approach 
 suggested in \cite {efr88,lyk93}, the use of which has reproduced the
 the experimental data about the proton spectrum in the deuteron
 stripping, allows us also to describe the inclusive pion spectrum 
 at all the values of $x_C$ rather well (fig.~\ref{fig:fig7}).
 However, the information
 contained in these both observables is redundant, since the main 
 ingredient by analyzing both reactions $D p\rightarrow p X$ and 
 $D p\rightarrow\pi X$ within the impulse approximation are the same
 deuteron properties. Therefore the calculation of the tensor analyzing
 power including the non-nucleon degrees of freedom in the fragmentation
 of deuteron to pions can give us new independent information about the 
 deuteron structure at small $N-N$ distances and its comparison 
 with data can be as a test of the used model of the modification 
 of the DWF.
 These results are presented in fig.\ref{fig:fig8} and show some improvement
 of description of the experimental data about $T_{20}$ \cite{afa98},
 especially at $x_{\cal C} > 1.3$,
 if we assume that the non-nucleon degrees of freedom contribute, mainly,
 to the ${\cal S}$- and ${\cal D}$-waves of the DWF.
 Of course, the performed inclusion of the non-nucleon degrees of freedom
 by analysis of $T_{20}$ is the approximate and can be considered as
 an indication of an important role of these degrees of freedom studying
 polarization phenomena in the discussed type of reactions. 
\end{enumerate}

\newpage

\end{document}